%% file: draft_D2Kpipienu_v2.0.tex
\documentclass[aps,prd,twocolumn,showpacs,amsmath,amssymb]{revtex4-2}
\usepackage{amsmath}
\usepackage{graphicx}
\usepackage{subfigure}
\usepackage{epstopdf}
\usepackage{color}
\usepackage{multirow}
\usepackage{setspace}
\usepackage{overpic}
\usepackage{amssymb}
\usepackage[title]{appendix}
\usepackage{makecell}
\usepackage[bookmarksnumbered, pdfstartview=FitH,colorlinks,urlcolor=blue, citecolor=blue,linkcolor=blue] {hyperref}
\usepackage{lineno}
\usepackage{bm}
\usepackage{float}
\usepackage{rotating}
\usepackage[utf8]{inputenc}
\usepackage[separate-uncertainty=true,table-align-uncertainty=true,]{siunitx}

\let\oldequation\equation
\let\oldendequation\endequation
\renewenvironment{equation}
 {\linenomathNonumbers\oldequation}
 {\oldendequation\endlinenomath}

\begin{document}
\title{\boldmath First Measurement of the Decay Dynamics  in the Semileptonic  Transition of the \\ $D^{+(0)}$ into the Axial-vector Meson $\bar K_1(1270)$}
\input{authorlist_2024-11-29}

\begin{abstract}
Using  $e^+e^-$ data taken at the center-of-mass energy of 3.773 GeV with the BESIII detector, corresponding to an integrated luminosity of 20.3 fb$^{-1}$, we report the first measurement of the decay dynamics of the semileptonic decays $D^{+(0)}\to K^-\pi^+\pi^{0(-)} e^+\nu_e$. The amplitude analysis gives the hadronic form factors of the semileptonic $D$ transitions into the axial-vector meson $\bar{K}_1(1270)$ to be $r_A=(-11.2\pm1.0_{\rm stat}\pm0.9_{\rm syst})\times10^{-2}$ and $r_V = (-4.3\pm 1.0_{\rm stat}\pm2.5_{\rm syst})\times 10^{-2}$. This is the first in the semileptonic decays of heavy mesons into axial-vector mesons. The angular analysis yields an up-down asymmetry $\mathcal{A}^\prime_{ud} = 0.01\pm0.11$, which is consistent with the Standard Model prediction. In addition, the branching fractions of $D^+\to \bar K_1(1270)^0 e^+\nu_e$ and 
$D^0\to K_1(1270)^- e^+\nu_e$ are determined with improved precision to be $(2.27\pm0.11_{\rm stat}\pm0.07_{\rm syst}\pm0.07_{\rm input})\times10^{-3}$ and $(1.02\pm0.06_{\rm stat}\pm0.06_{\rm syst}\pm0.03_{\rm input})\times10^{-3}$, respectively. No significant signals of $D^+\to \bar K_1(1400)^0 e^+\nu_e$ and 
$D^0\to K_1(1400)^- e^+\nu_e$ are observed and their branching fraction upper limits are set as $1.4\times10^{-4}$ and $0.7\times10^{-4}$ at 90\% confidence level, respectively.

\end{abstract}

\maketitle

The partial decay rates of semileptonic (SL) $D$ decays are usually decomposed as functions of the quark mixing matrix element $|V_{cs}|$ or $|V_{cd}|$~\cite{Cabibbo:1963yz,Kobayashi:1973fv} and the hadronic form factors (FFs) describing strong interaction effects binding quarks into hadrons. Over the past few decades, the hadronic FFs of the SL $D$ decays into $S$-wave states (pseudoscalar or vector mesons) have been precisely and extensively studied in theory~\cite{FlavourLatticeAveragingGroupFLAG:2024oxs} and experiment~\cite{HFLAV:2022esi,Ke:2023qzc}. However, there is scant information regarding the SL decay of $D$ into $P$-wave states. As a Cabbibo-favored transition into the lowest lying axial-vector meson octet, the $D\to K_1(1270)$ transition is expcted to be the most promising channel with significantly higher statistics. There have been some model predictions on its hadronic FFs~\cite{Momeni:2019uag,Khosravi:2008jw,Momeni:2022gqb,Verma:2011yw,Cheng:2003sm}, but no any experimental information yet. 


In theory, the physical mass eigenstates of the strange axial-vector mesons, $K_1(1270)$ and $K_1(1400)$, are mixture of the $^1P_1$ and $^3P_1$ states with a mixing angle $\theta_{K_1}$.
Some theoretical calculations of the hadronic FFs of $D\to \bar{K}_1(1270)$ have come from light-cone QCD sum rules~(LCSR)~\cite{Momeni:2019uag}, three-point QCD sum rules~(3PSR)~\cite{Khosravi:2008jw}, Ads/QCD~\cite{Momeni:2022gqb}, and covariant light front approach~(LFQM)~\cite{Verma:2011yw,Cheng:2003sm}. The predicted FFs are in wide range due to being sensitive to both the theoretical approaches and $\theta_{K_1}$. However, the value of $\theta_{K_1}$ is still very controversial in various phenomenological analyses of different processes~\cite{Suzuki:1993yc,Divotgey:2013jba,Hatanaka:2008xj,Shi:2023kiy,Blundell:1995au,Cheng:2011pb,Tayduganov:2011ui,Lipkin:1977uy,Burakovsky:1997dd}. 
Experimental measurements of the hadronic FFs of $D\to K_1(1270)$ are crucial to test different theoretical calculations, and thereby restrict the $\theta_{K_1}$.
A verified $\theta_{K_1}$ value is key to guide the theoretical calculations of different decays of $\tau$~\cite{Suzuki:1993yc}, $B$~\cite{Hatanaka:2008xj,Cheng:2004yj}, and $D$~\cite{Cheng:2003bn,Guo:2018orw} particles into strange axial-vector mesons.


Additionally, Refs.~\cite{Wang:2019wee,Bian:2021gwf} state that the analysis of $B\to K_1(1270)\gamma$ when combined with $D^{0(+)}\to \bar K_{1}(1270) \ell^+ \nu_\ell$ offers a potential way to extract photon polarization in $b\to s\gamma$ transitions without considerable theoretical ambiguity. The improved knowledge of photon polarization in $b\to s\gamma$ transitions is powerful to probe right-handed couplings in new physics~\cite{Atwood:1997zr,Becirevic:2012dx,Paul:2016urs}. Previously, a significant photon polarization was observed in the $B^{+}\to K_1(1270)^+(\to K^{+}\pi^{-}\pi^+)\gamma$ decay by measuring its up-down asymmetry $\mathcal{A}_{ud}=f_{h}\lambda_\gamma$~\cite{LHCb:2014vnw}. Therefore, determination of the up-down asymmetry $\mathcal{A}^\prime_{ud}$ in the SL decays $D^{+(0)}\to \bar{K}_{1}(1270)^{0(-)}(\to K^-\pi^+\pi^{0(-)})e^+\nu_e$, is highly desired to quantify the hadronic effects of $K_{1}(1270)^-\to K^-\pi^+\pi^-$ with $f_h=\frac{3}{4}\mathcal{A}^\prime_{ud}$.


Previously, only CLEO and BESIII reported the branching fraction~(BF) measurements of $D^{+(0)}\to \bar{K}_{1}(1270)^{0(-)}e^+\nu_e$~\cite{CLEO:2007oer,BESIII:2019eao,BESIII:2021uqr,BESIII:2024ieo}. This Letter reports the first determinations of the hadronic FFs of $D\to K_1(1270)$ ($r_A$ and $r_V$) and up-down asymmetry ($\mathcal{A}^\prime_{ud}$), using 20.3 fb$^{-1}$ of $e^+e^-$ data~\cite{BESIII:2024lbn} collected by BESIII at $\sqrt{s}=$ 3.773 GeV.
Improved measurements of the BFs of $D\to K_1(1270)e^+\nu_e$ and the first search for $D\to K_1(1400)e^+\nu_e$ are also presented~\cite{cc_ref}.

Details about the design and performance of the BESIII detector are given in Refs.~\cite{BESIII:2009fln,Huang:2022wuo}. The inclusive Monte Carlo~(MC) samples, described in Refs.~\cite{GEANT4:2002zbu,Jadach:1999vf,Jadach:2000ir,Lange:2001uf,Ping:2008zz,Richter-Was:1992hxq}, are used to model the background in this analysis. The signal MC samples of the $D^{+(0)}\to K^-\pi^+\pi^{0(-)} e^+\nu_e$ decay are generated based on the amplitude analysis result obtained in this work.


At $\sqrt{s}=3.773$ GeV, the $D$ and $\bar{D}$ mesons are produced in pairs without accompanying particles in the final state; this allows us to study SL $D$ decays with the double-tag~(DT) method~\cite{BESIII:2023exq}. The single-tag~(ST) $\bar{D}$ mesons are reconstructed with the hadronic decay modes $\bar{D}^0\to K^+\pi^-,K^+\pi^-\pi^0$, and $K^+\pi^-\pi^-\pi^+$; $D^-\to K^+\pi^-\pi^-,K_S^0\pi^-,K^+\pi^-\pi^-\pi^0,K_S^0\pi^-\pi^0,K_S^0\pi^+\pi^-\pi^-$, and $K^+K^-\pi^-$. In the presence of the ST $\bar{D}$ mesons, candidates for the signal decays $D^{+(0)}\to K^-\pi^+\pi^{0(-)} e^+\nu_e$ are selected to form DT events. The BF of the signal decay is determined by
\begin{equation}
\label{eq:BF}
	\mathcal{B}_{\rm sig} = \frac{N^{\rm sig}_{\rm DT}}{N^{\rm tot}_{\rm ST}\bar{\epsilon}_{\rm sig}},
\end{equation}
where $N^{\rm tot}_{\rm ST}=\sum_{i}{N^i_{\rm ST}}$ and $N^{\rm sig}_{\rm DT}$ are the total ST and DT yields after summing over all tag modes; $\bar{\epsilon}_{\rm sig}=\sum_{i}\frac{N^i_{\rm ST}}{N^{\rm tot}_{\rm ST}}\frac{\epsilon^i_{\rm DT}}{\epsilon^i_{\rm ST}}$ is the averaged signal efficiency of selecting $D^{+(0)}\to K^-\pi^+\pi^{0(-)} e^+\nu_e$ in the presence of the ST $\bar{D}$ mesons, where $\epsilon^i_{\rm ST}$ and $\epsilon^i_{\rm DT}$ are the ST and DT efficiencies for the $i$-th tag mode, respectively.

The selection criteria for $\pi^\pm,K^\pm,K_S^0,\gamma$, and $\pi^0$ candidates are the same as Ref.~\cite{BESIII:2023exq}. Two kinematic variables, the energy difference $\Delta E \equiv E_{\bar{D}}-E_{\rm beam}$ and the beam-constrained mass $M_{\rm BC}\equiv \sqrt{E_{\rm beam}^2-|\vec{p}_{ \bar{D}}|^2}$, are used to distinguish the ST $\bar{D}$ mesons from the combinatorial background, where $E_{\rm beam}$ is the beam energy and $(E_{\bar{D}}, \vec{p}_{ \bar{D}})$ is the four-momentum of ST $\bar{D}$ in the $e^+e^-$ rest frame. The combination with the smallest $|\Delta E|$ is chosen if there are multiple combinations in the event.

The ST candidates are required to satisfy the tag mode requirements of $\Delta E$, corresponding to about $\pm 3.5\sigma$ around the fitted peaks. 
The ST yields in data for each tag mode are extracted by fitting individual $M_{\rm BC}$ distributions. Candidates with $M_{\rm BC}\in [1.859,1.873]$~GeV/$c^2$ for $\bar D^0$ and $M_{\rm BC}\in [1.863,1.877]$~GeV/$c^2$ for $D^-$ are retained for further analysis. The $\Delta E$ requirements, the ST yields in data, and the ST efficiencies for different tag modes are provided in Ref.~\cite{supplemental_material}. Summing over all tag modes gives $N^{\rm tot}_{\rm ST}$ to be $(10677.9\pm3.8)\times10^3$ for $D^-$ and $(16146.2\pm4.6)\times10^3$ for $\bar{D}^0$.


In the presence of the tagged $\bar{D}$, candidates for  $D^{+(0)}\to K^-\pi^+\pi^{0(-)} e^+\nu_e$ are selected from the residual tracks and showers. The selection criteria of $K^-,\pi^\pm$, and $\pi^0$ candidates are the same as those used in the ST selection. The $e^+$ candidates are selected by performing particle identification~(PID) based on the specific ionization energy loss ${\rm d}E/{\rm d}x$, time of flight, and electromagnetic calorimeter~(EMC) information. Confidence levels~($CL$) for the positron, pion, and kaon hypotheses, denoted as $CL_{e,\pi,K}$, are calculated. Charged tracks are assigned as $e^+$ candidates if $CL_{e}>0.001$ and $CL_{e}/(CL_{e}+CL_{\pi}+CL_{K})>0.8$. To further distinguish the positron candidates from hadrons, the $e^+$ candidates are required to satisfy $E/p-0.05\times\chi^2_{e-{{\rm d}E/{\rm d}x}}>0.53$ and $0.60$ for $D^+$ and $D^0$ channels, respectively. Here, $E$ is the deposited energy in the EMC, $p$ is the momentum measured by the drift chamber~(MDC), and $\chi^2_{e-{\rm d}E/{\rm d}x}$ is the $\chi^2$ with the positron hypothesis based on the ${\rm d}E/{\rm d}x$ information. To partially recover the energy loss due to final state radiation~(FSR) for positrons, the neutral showers within 5$^\circ$ of the initial positron direction are merged to the four-momentum of the positron measured by the MDC. No additional charged tracks are allowed in the event to suppress the hadronic background.

Since the neutrino cannot be detected by the BESIII detector, its four-momentum $(E_{\rm miss},\vec{p}_{\rm miss})$ is obtained by calculating the missing energy and momentum, defined as $E_{\rm miss}\equiv E_{\rm beam}-\sum_{j}E_j$ and $\vec{p}_{\rm miss}\equiv \vec{p}_{D}-\sum_j\vec{p}_{j}$. Here, the index $j$ sums over the $K^-,\pi^+,\pi^{0(-)}$, and $e^+$ of the signal candidates and $(E_j,\vec{p}_{j})$ is the four-momentum of the $j$-th particle. A kinematic variable $U_{\rm miss}\equiv E_{\rm miss} - |\vec{p}_{\rm miss}|$ is defined to extract the signal yield. To further improve the resolutions of the above variables, the momentum of the $D$ meson is constrained as $\vec{p}_{D}=-\hat{p}_{\bar{D}}\sqrt{E^2_{\rm beam}-m^2_{\bar{D}}}$, where $\hat{p}_{\bar{D}}$ is the unit vector in the momentum direction of the ST $\bar{D}$ meson and $m_{\bar{D}}$ is the known $\bar{D}$ mass~\cite{ParticleDataGroup:2024cfk}.

Further requirements are applied to veto the background following Refs.~\cite{BESIII:2019eao,BESIII:2021uqr}. For the $D^+$ channel, the momentum of the $\pi^0$ candidate must be greater than 0.2~GeV/$c$ to suppress the fake $\pi^0$ candidates; the invariant mass of $K^-\pi^+\pi^0e^+$ is required to be less than 1.78~GeV/$c^2$ to reject the hadronic decay $D^+\to K^- \pi^+ \pi^0 \pi^+$; to suppress the background from $D^+\to K^-\pi^+ e^+\nu_e$ with a fake $\pi^0$, the $U^\prime_{\rm miss}$ variable is calculated by ignoring the $\pi^0$, and events with $U^\prime_{\rm miss}<0.03$ GeV are rejected. For the $D^0$ channel, the invariant mass of $K^-\pi^+\pi^-\pi^+_{e\to\pi}$ is required to be less than 1.80~GeV/$c^2$ to reject the hadronic decay $D^0\to K^- \pi^+ \pi^- \pi^+$, where $\pi^+_{e\to\pi}$ is obtained by replacing the energy of the $e^+$ track with the pion mass; the opening angle between $e^+$ and $\pi^-$ is required to satisfy $\cos\theta_{e^+\pi^-}<0.93$ to suppress the background from $D^0\to K^-\pi^+\pi^0(\pi^0),\pi^0\to e^+e ^-\gamma$ with $e^-$ mis-identified as $\pi^-$; the background $D^0\to K^-\pi^+\pi^-\pi^+ \pi^0$ is suppressed by requiring $\cos(\nu_e,\gamma)<0.78$, where $\gamma$ is the most energetic extra photon; the background from $D^0\to K^-\pi^0 e^+\nu_e, \pi^0\to e^+e^-\gamma$ with $e^+e^-$ mis-identified as $\pi^+\pi^-$ is rejected by requiring the $\pi^+\pi^-$ invariant mass to be less than 0.31~GeV/$c^2$; events with $D^-\to K^+\pi^-\pi^-$ versus $D^+\to \pi^0X$ can be potentially reconstructed as the tag mode $\bar{D}^0\to K^+\pi^-\pi^0$, and are suppressed by requiring $|\Delta E[(K^+\pi^-)_{\rm tag}\pi^-_{\rm sig}]|>7$ MeV.

\begin{figure}[htbp]
	\centering
	\vspace{-0.0cm}
	\setlength{\abovecaptionskip}{0cm}
	\includegraphics[width=8.7cm]{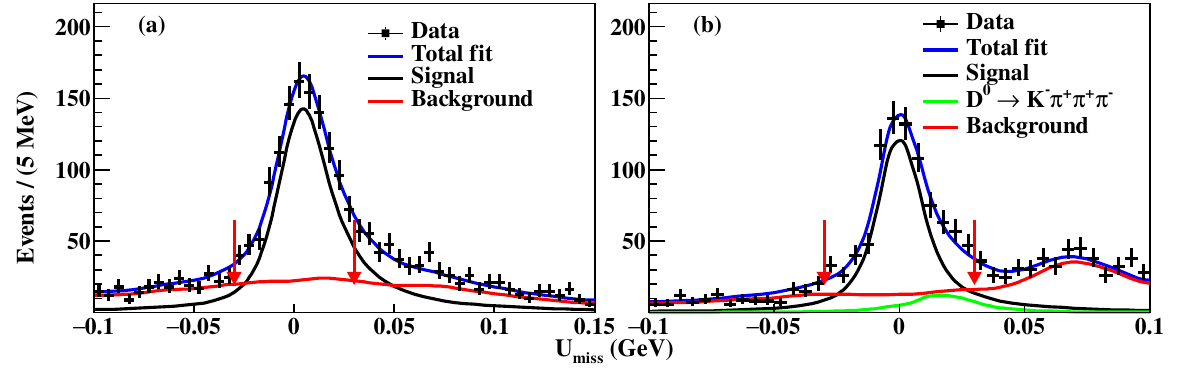}
	\caption{Fits to the $U_{\rm miss}$ distributions of the (a) $D^+$ and (b) $D^0$ channels. The points with error bars are data. The blue solid lines denote the total fits. The black, green, and red lines show the signal, peaking background, and non-peaking background contributions, respectively. The pairs of red arrows indicate the requirement $|U_{\rm miss}|<0.03$ GeV.}
	\label{fig:fit_umiss}
\end{figure}

To extract the signal yields of $D^{+(0)}\to K^-\pi^+\pi^{0(-)} e^+\nu_e$, unbinned maximum likelihood fits are performed on the $U_{\rm miss}$ distributions of the accepted candidates, as shown in Fig.~\ref{fig:fit_umiss}. In the fit, the signal is modeled by the MC-simulated shape convolved with a Gaussian function with free parameters, and the background shape is derived from the inclusive MC sample. For the $D^0$ channel, the peaking background $D^0\to K^-\pi^+\pi^-\pi^+$ is modeled based on the amplitude analysis result~\cite{BESIII:2017jyh} with fixed yield according to the simulated sample, while the yield of the other background components is allowed to float. There is no significant peaking background for the $D^+$ channel and the yield of all the background contributions is allowed to float. The signal efficiencies are estimated based on the signal MC samples generated with the amplitude analysis result. The fitted signal yields from the data, the averaged signal efficiencies and the obtained BFs are summarized in Table~\ref{tab:Bfs}. 


\begin{table}[htbp]
	\vspace{-0.9cm}
	\setlength{\abovecaptionskip}{0cm}
	\centering
	\setlength\tabcolsep{5pt}
	\caption{Fitted signal yields in data ($N^{\rm sig}_{\rm DT}$), averaged signal efficiencies ($\bar{\epsilon}_{\rm sig}$), and obtained BFs ($\mathcal{B}_{\rm sig}$), where the first uncertainties are statistical and the second systematic.}
	
	\begin{tabular}{cccc}
		\hline\hline
		Channel& $N^{\rm sig}_{\rm DT}$ &$\bar{\epsilon}_{\rm sig}$~(\%)&$\mathcal{B}_{\rm sig}~(\times10^{-3})$ \\
		\hline
		$D^+$ & $1270\pm 56$ &$9.45\pm0.02$ &$1.27\pm0.06\pm0.04$ \\
		$D^0$ & $731\pm35$ &$14.12\pm0.03$&$0.32\pm0.02\pm0.02$\\
		\hline\hline
	\end{tabular}
	\label{tab:Bfs}
\end{table}

\begin{figure*}[htbp]
	\centering
	\setlength{\abovecaptionskip}{0.0cm}
	\includegraphics[width=18.5cm]{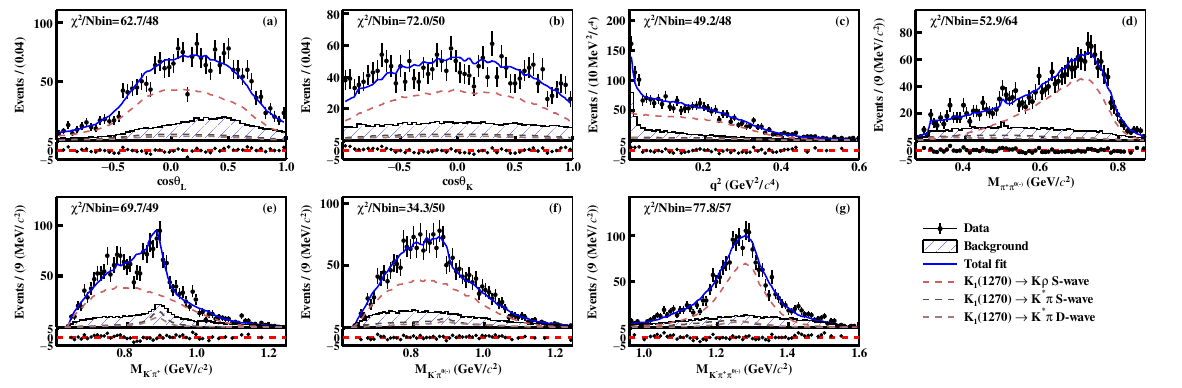}
	\caption{Projections of the amplitude analysis result on (a) $\cos\theta_{L}$, (b) $\cos\theta_{K}$, (c) $q^2$, (d) $M_{\pi^+\pi^{0(-)}}$, (e) $M_{K^-\pi^+}$, (f) $M_{K^-\pi^{0(-)}}$, and (g) $M_{K^-\pi^+\pi^{0(-)}}$. The dots with error bars are data, the blue solid curves are the total fit results, the dashed curves represent the various components, and the blue hatched histograms are the simulated backgrounds derived from the inclusive MC sample. The pull distributions, defined as $\chi=(N_{\rm data}-N_{\rm fit})/\sqrt{N_{\rm fit}}$, are shown in the lower plots, where $N_{\rm data}$ and $N_{\rm fit}$ are the data and fit projection numbers of events for each bin, respectively.}
	\label{fig:fit_projection}
\end{figure*}

Events satisfying $|U_{\rm miss}|<0.03$ GeV are kept for the amplitude analysis. In total 1226 and 882 data candidates for the $D^+$ and $D^0$ channels survive with background fractions of $(21.9\pm1.4)\%$ and $(26.7\pm1.4)\%$, respectively. In the amplitude analysis, the covariant tensor amplitude is constructed as
\begin{equation}
\begin{aligned}
\mathcal{M} & =(V-A)^{\mu\eta}\cdot
[\sum_{\lambda_{W}}\epsilon^{*}(\lambda_{W})_{\mu}\epsilon(\lambda_{W})_{\rho}] \cdot\\ 
&
[\sum_{\lambda_{K_1}}\epsilon^{*}(\lambda_{K_1})_{\eta}\epsilon(\lambda_{K_1})_{\sigma}]\cdot \mathcal{R}_{\bar{K}_1}\cdot J^{\sigma}\cdot\bar{u}_\nu\gamma^{\rho}(1-\gamma_5)v_l.
\end{aligned}
\end{equation}
Here, $(V-A)^{\mu\eta}\epsilon^{*}(\lambda_{K_1})_{\eta}$ is the current for $D\to \bar{K}_1 W^*$ following the convention in Ref.~\cite{Bian:2021gwf}, written as
\begin{equation}
\begin{aligned}
V^{\mu\eta} \epsilon^{*}(\lambda_{K_1})_{\eta}  =&
-(m_D-M_{K_1}) V_1(q^2)\epsilon^{*\mu}(\lambda_{K_1}) \\
&+V_2(q^2) \left(\frac{q\cdot\epsilon^{*}(\lambda_{K_1})}{m_D-M_{K_1}}\right)(p_D+p_{K_1})^\mu,\\
A^{\mu\eta} \epsilon^{*}(\lambda_{K_1})_{\eta} =&-\frac{2 i A(q^2)}{m_D-M_{K_1}} \epsilon^{\mu \epsilon^*(\lambda_{K_1}) p^{D}p^{K_1}}. \\
\end{aligned}
\end{equation}
Both vector  and axial-vector FFs take single pole form, written as  $V_{1,2}(q^2)= \frac{V_{1,2}(0)}{1-q^2/m_V^2}$ and $A(q^2)=\frac{A(0)}{1-q^2/m_A^2}$ with $q^\mu=p^\mu_{D}-p_{K_1}^\mu$; 
$\epsilon(\lambda_{W})$ and $\epsilon(\lambda_{K_1})$ are the polarization vectors of the W-boson and the $\bar{K}_1$ meson, respectively; $\mathcal{R}_{\bar{K}_1}=\frac{1}{s-m_0^2+im_0\Gamma_0}$ is the Breit-Wigner function of $\bar{K}_1$ with mass $m_0$ and width $\Gamma_0$; $J^\sigma$ is the hadronic current of $\bar{K}_1\to K^-\pi^+\pi^{0(-)}$; $W^*\to e^+\nu_e$ is described with $\bar{u}_\nu\gamma^{\rho}(1-\gamma_5)v_l$. See more details about the amplitude formula in Ref.~\cite{supplemental_material}. For the charge conjugate decay modes, the three-momenta of the final states from $\bar{D}$ decay are inverted  to incorporate them with the $D$ by assuming charge-parity conservation.

The log-likelihood function for this unbinned maximum likelihood fit is constructed as
\begin{equation}
\begin{aligned}
\ln \mathcal{L} &=\sum^{N_{\rm data}}_{k} \ln[(1-\omega_{\rm bkg})\frac{|\mathcal{M}(p^k_{j})|^2}{\int\epsilon(p_{j})|\mathcal{M}(p_j)|^2 R_{5}(p_j){\rm d}p_j}\\
&+\omega_{\rm bkg}\frac{{B}_{\epsilon}(p^k_{j})}{\int\epsilon(p_{j}){B}_{\epsilon}(p_j) R_{5}(p_j){\rm d}p_j}],
\end{aligned}
\end{equation}
where $N_{\rm data}$ is the number of events in the data sample; $\omega_{\rm bkg}$ is the background fraction; $p_{j}$ is the four-momentum of final states; ${B}_{\epsilon}$ is the background distribution corrected by acceptance. The integration over five-body phase space $R_{5}$ is calculated numerically~\cite{BESIII:2015hty} based on massive MC samples. A simultaneous fit is performed to $D^+$ and $D^0$ channels by summing over their log-likelihood functions. The isospin relationships are imposed to constrain the FFs in the SL decays and complex coupling coefficients in the hadronic current $J^\sigma$ as shown in Ref.~\cite{supplemental_material}. Here, $V_{1}(0)$ is fixed to be 1 with the ratio $r_V=\frac{V_{2}(0)}{V_{1}(0)}$ and $r_A=\frac{A(0)}{V_{1}(0)}$ allowed to float. The mass and width of $K_1(1270)$ are free in the fit.

Several potential sub-structures in $\bar{K}_1(1270)\to K^-\pi^+\pi^{0(-)}$, including $\rho,\bar{K}^*(892),\omega$, and $(\bar{K}\pi)_{S-{\rm wave}}$, are considered in the significance test. The contribution from $\bar{K}_1(1400)\to \bar{K}^*(892)\pi$ is also tested, which yields statistical significance less than 3$\sigma$. The nominal solution is determined to include $\bar{K}_1(1270)\to \rho K^-$ ($S$-wave) and $\bar{K}_1(1270)\to \pi \bar{K}^*(892)$ ($S$- and $D$-waves) with statistical significance greater than 5$\sigma$. Figure~\ref{fig:fit_projection} shows the projections of the nominal fit result. The kinematic variable $\cos\theta_{L}$ is the angle between the vector opposite to the flight direction of $D$ and $e^+$ in the W-boson rest frame;
$\cos\theta_{K}$ is the angle between the vector opposite to the flight direction of $D$ and the normal to the $K_1$ decay plane $\vec{n}=\vec{p}_{{\pi}^{+}}\times\vec{p}_{{\pi}^{0(-)}}$ in the $\bar{K}_1$ rest frame. The projections on helicity angles in the secondary decay $\bar{K}_1(1270)\to K^-\pi^+\pi^{0(-)}$ are presented in Appendix~\ref{sec:proj_hel_angle}. The fitted parameters and the fit fractions are summarized in Table~\ref{tab:fited_pars}. Here, the fit fraction $f_{i}$ of the $i$-th component is determined with
\begin{equation}
	f_{i} = \int{|\mathcal{M}_i|^2 R_{5}(p_j)dp_j} \bigg/ \int{|\mathcal{M}|^2 R_{5}(p_j)dp_j},
\end{equation}
where $\mathcal{M}_i$ is the amplitude with the $i$-th component's contribution only. The BF ratio of $\frac{\mathcal{B}(K_1(1270)\to K^*\pi)}{\mathcal{B}(K_1(1270)\to K\rho)}$ is calculated to be $(20.3\pm 2.1_{\rm stat}\pm 8.7_{\rm syst})\%$, which is consistent with those derived from Refs.~\cite{Belle:2010wrf,LHCb:2017swu,LHCb:2024cwp} but disfavors those obtained in Refs.~\cite{BESIII:2021qfo,LHCb:2018mzv}. Furthermore, with $\bar{K}_1(1400)\to \bar{K}^*(892)\pi$ involved in fit, the upper limit of corresponding fit fraction is set to be 9\% for $D^0$ and 5\% for $D^+$ at 90\% $CL$.

By applying the isospin relationship and ignoring the insignificant contribution from $K_1(1270)\to K^*_0(1430)\pi$, the BFs of $\mathcal{B}(\bar{K}_1(1270)^0\to K^-\pi^+\pi^0)=\mathcal{B}_{\rm{3body}}\cdot\frac{3+4\alpha}{9(1+\alpha)}$ and $\mathcal{B}(K_1(1270)^-\to K^-\pi^+\pi^-)=\mathcal{B}_{\rm{3body}}\cdot\frac{6+4\alpha}{9(1+\alpha)}$ are calculated to be $(56.0\pm1.7)\%$ and $(31.3\pm0.9)\%$, respectively. Here, $\mathcal{B}_{\rm{3body}}=1-\mathcal{B}(K_1(1270)\to K\omega)$ is the BF of the three-body decay of $K_1(1270)$ and $\alpha=\frac{\mathcal{B}(K_1(1270)\to K^*\pi)}{\mathcal{B}(K_1(1270)\to K\rho)}$. We further determine $\mathcal{B}(D^+\to \bar{K}^0_1(1270)e^+\nu_e)=(2.27\pm0.11_{\rm stat}\pm0.07_{\rm syst}\pm0.07_{\rm input})\times10^{-3}$ and $\mathcal{B}(D^0\to {K}^-_1(1270)e^+\nu_e)=(1.02\pm0.06_{\rm stat}\pm0.06_{\rm syst}\pm0.03_{\rm input})\times10^{-3}$. Our BFs of $D\to K_1(1270)e^+\nu_e$ are consistent with the previous measurements~\cite{CLEO:2007oer,BESIII:2019eao,BESIII:2021uqr,BESIII:2024ieo} but with improved precision. The upper limits of $\mathcal{B}(D^+\to \bar{K}^0_1(1400)e^+\nu_e)$ and $\mathcal{B}(D^0\to {K}^-_1(1400)e^+\nu_e)$ are set to be $1.4\times10^{-4}$ and $0.7\times10^{-4}$ at 90\% $CL$, respectively.

\begin{table}[htbp]
	\centering
	\setlength\tabcolsep{20pt}
	\caption{Fitted parameters and fit fractions, where the first uncertainties are statistical and the second systematic.}
	\begin{tabular}{lc}
		\hline
		\hline
		Variable & Value \\ \hline
		$r_{A}~(\times10^{-2})$ & $-11.2\pm1.0\pm0.9$  \\ 
		$r_{V}~(\times10^{-2})$ & $-4.3\pm1.0\pm2.5$\\
		$f^{D^+}_{\rho K^-}$~(\%) & $79.3\pm2.0\pm25.7$\\
		$f^{D^+}_{\pi \bar{K}^*(892)}$~(\%) & $10.9\pm1.2\pm3.0$\\
		$f^{D^+}_{\bar{K}_1(1400)}$~(\%) & $<5$\\
		$f^{D^0}_{\rho K^-}$~(\%) & $71.8\pm2.3\pm23.9$\\
		$f^{D^0}_{\pi \bar{K}^*(892)}$~(\%) & $19.5\pm1.9\pm5.2$\\
		$f^{D^0}_{\bar{K}_1(1400)}$~(\%) & $<9$\\
		$m_{K_1(1270)}$ (MeV/$c^2$)  & $1271\pm3\pm7$  \\ 
		$\Gamma_{K_1(1270)}$ (MeV)  & $168\pm10\pm20$  \\ 
		\hline
		\hline
	\end{tabular}
	\label{tab:fited_pars}
\end{table}

An additional binned two-dimensional $\chi^2$-fit is performed on the $\cos\theta_{L}$ versus $\cos\theta_K$ distributions of the $D^+$ and $D^0$ channels simultaneously, to extract the up-down asymmetry $\mathcal{A}^\prime_{ud}$ in the $D\to \bar{K}_1(1270) e^+\nu_e$ decay. This asymmetry relates to the photon polarization $\lambda_\gamma$ in $b\to s\gamma$ via $\lambda_{\gamma}=\frac{4}{3}\frac{\mathcal{A}_{ud}}{\mathcal{A}^\prime_{ud}}$~\cite{Bian:2021gwf}, where $\mathcal{A}_{ud}=0.069\pm0.017$ is the up-down asymmetry in $B^{+}\to K_1(1270)^+(\to K^{+}\pi^{-}\pi^+)\gamma$~\cite{LHCb:2014vnw}. The probability density function $f(\cos\theta_{L},\cos\theta_K; \mathcal{A}^\prime_{ud}, d_{+}, d_{-})$ is cited from Ref.~\cite{Fan:2021mwp}, where $d_{\lambda_{W}}=|c_{\lambda_{w}}|^2/|c_{0}|^2$ is the contribution ratio of $\lambda_{w}=\pm$ to $\lambda_{w}=0$. To keep consistent with the convention in Ref.~\cite{LHCb:2014vnw}, the direction normal to the $K_1$ decay plane is defined as $\vec{n}=\vec{p}_{\pi,{\rm slow}}\times\vec{p}_{\pi,{\rm fast}}$ instead, and additional mass window $M_{\bar{K}\pi\pi}\in[1.1,1.3]$ GeV/$c^2$ is required. The construction of the $\chi^2$ function and the fit projections are described in Appendix~\ref{sec:angular_fit}. The fit gives $\mathcal{A}^\prime_{ud}=0.01\pm0.11$, which is consistent with the SM prediction $(0.092\pm0.022)$. Additionally, the fraction of $\bar{K}_1(1270)$ longitudinal polarization $F_L=|c_0|^2/(|c_0|^2+|c_+|^2+|c_-|^2)=1/(1+d_{+}+d_{-})$ is determined to be $0.50\pm0.04$ with statistical uncertainty only, which is consistent with Ref.~\cite{BESIII:2021uqr} but with four times higher precision. Here, the potential systematic uncertainties are negligible.

The systematic uncertainties in the BF measurements are discussed below. The uncertainty associated with the ST yield $N^{\rm tot}_{\rm ST}$ is estimated to be 0.3\%~\cite{BESIII:2024zfv}. The efficiencies of tracking, PID, and $\pi^0$ reconstruction are studied with $e^+e^-\to\gamma e^+e^-$ and $D\bar{D}$ hadronic decays. The uncertainties are assigned to be 0.5\% for tracking or PID for each charged track and 1.0\% for each $\pi^0$. The uncertainties associated with various cuts are studied with the control samples $D^{+(0)}\to K^0_S\pi^{0(-)}e^+\nu_e$, which are estimated to be $2.0\%$ for $D^+$ and $4.6\%$ for $D^0$, respectively. The uncertainties of the $U_{\rm miss}$ fits are estimated to be 0.2\% for both $D^+$ and $D^0$ channels by changing the fit range and background shapes, as well as varying the peaking background yield. The uncertainties due to the MC model are estimated by varying the parameters in the amplitude model when generating the MC samples. The relative signal efficiency variations are 0.8\% and 0.9\% for $D^+$ and $D^0$, respectively. The uncertainty of FSR recovery is assigned to be 0.3\%~\cite{BESIII:2015tql}. Combining all individual contributions in quadrature, the overall uncertainties are 3.3\% and 5.5\% for $D^+$ and $D^0$, respectively.

The systematic uncertainties of the amplitude analysis are given below. The uncertainty due to non significant components is estimated by including them into the solution separately. The uncertainty of the radius parameter in the Blatt-Weisskopf factor~\cite{ParticleDataGroup:2024cfk} is estimated by varying it from 3 GeV$^{-1}$ to 2 and 4 GeV$^{-1}$. The uncertainty due to the lineshape of $\bar{K}_1(1270)$ is estimated by using the alternative energy-dependent width $\Gamma(s)\propto\frac{1}{\sqrt{s}}\int |M_{R\to abc}|^2 {\rm d}\Phi_{3}$~\cite{LHCb:2017swu}. The uncertainty associated with the background level is estimated by varying it within its uncertainty. The uncertainty due to parametrization of FFs is estimated by applying alternative modified pole model and parameter series expansion model following $D\to \bar{K}\ell^+\nu_\ell$~\cite{CLEO:2009svp,BESIII:2024slx}. The maximum changes on the amplitude fit results are assigned as the corresponding uncertainties. The uncertainty of fit bias, which is caused by the resolution effects as well as imperfect background modeling, is studied via 200 toy MC samples with background included. Amplitude fits are repeated to each sample, and the pull distributions of the fit results are fitted by a Gaussian function. The obtained mean values are assigned as this uncertainty. Detailed systematic uncertainties are given in Ref.~\cite{supplemental_material}.

\begin{figure}[htbp]
	\vspace{-0.4cm}
	\setlength{\belowcaptionskip}{0cm} 
	\centering
	\setlength{\abovecaptionskip}{-0.3cm}
	\includegraphics[width=8.5cm]{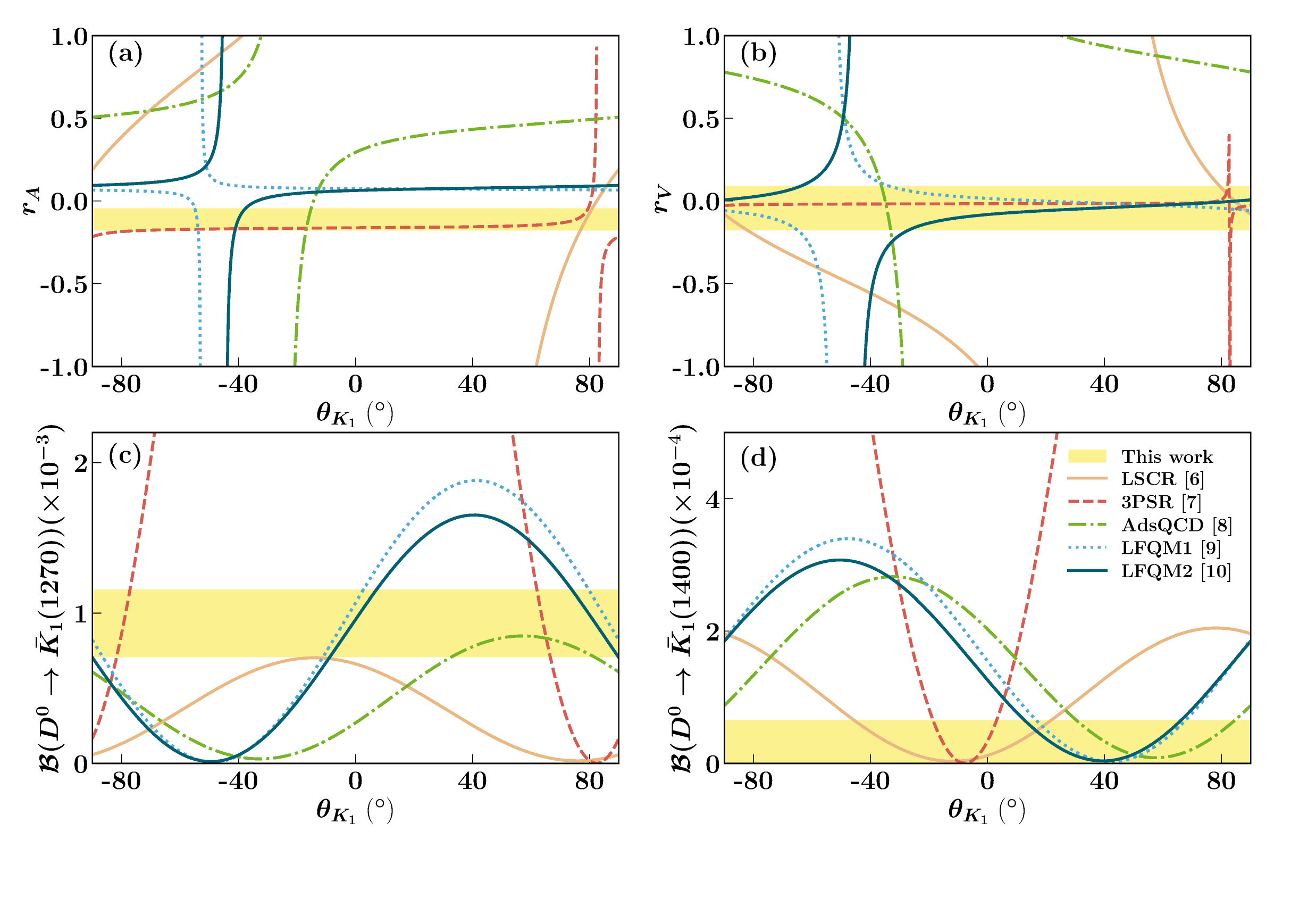}
	\caption{Comparisons of (a)~$r_A$, (b)~$r_V$, (c) $\mathcal{B}(D^0 \to K(1270)^-e^+\nu_e)$, and (d) $\mathcal{B}(D^0 \to K(1400)^- e^+\nu_e)$ measured in this work and predicted by various theoretical approaches as a function of $\theta_{K_1}$. The yellow bands show $\pm 5\sigma$ limit for (a), (b), (c), and the upper limit at 90\% $CL$ for (d).}
	\label{fig:exp_th_compare}
\end{figure}

In summary, we have made the first measurements of the hadronic FFs of $D\to K_1(1270)$, which are the first in the SL transitions of heavy mesons into axial-vector mesons. We have also presented the improved measurements of the BFs of $D\to K_1(1270)e^+\nu_e$ and the first search for $D\to K_1(1400)e^+\nu_e$. Figure~\ref{fig:exp_th_compare} shows a comparisons of the measured FFs and BFs with different theoretical predictions~\cite{Momeni:2019uag,Momeni:2022gqb,Khosravi:2008jw,Cheng:2003sm,Verma:2011yw} as a function of $\theta_{K_1}$. Our $r_A$, $r_V$, and $\mathcal{B}(D\to K_1(1270) e^+\nu_e)$ are consistent with the 3PSR predictions~\cite{Khosravi:2008jw} with $\theta_{K_1}\in(61,67)^\circ$,
and disfavor all other theoretical calculations by more than $5\sigma$. However, our upper limits on $\mathcal{B}(D\to K_1(1400)e^+\nu_e)$ disfavor the corresponding 3PSR predictions. More universal theoretical calculations for all four variables are still desired. Additionally, the up-down asymmetry $\mathcal{A}^\prime_{ud}$ is extracted for the first time and no new physics effect is found with the current statistics. Forthcoming larger $B\to K_1(1270)\gamma$~\cite{Belle-II:2022cgf,LHCb:2023hlw} and $D\to K_1(1270) e^+\nu_e$~\cite{Cheng:2022tog} samples are expected to provide more effective restriction on the right-handed couplings in new physics models.


The BESIII Collaboration thanks the staff of BEPCII and the IHEP computing center for their strong support. This work is supported in part by National Key R\&D Program of China under Contracts Nos. 2023YFA1606000; National Natural Science Foundation of China (NSFC) under Contracts Nos. 11635010, 11735014, 11935015, 11935016, 11935018, 12025502, 12035009, 12035013, 12061131003, 12192260, 12192261, 12192262, 12192263, 12192264, 12192265, 12221005, 12225509, 12235017, 12361141819; the Chinese Academy of Sciences (CAS) Large-Scale Scientific Facility Program; the CAS Center for Excellence in Particle Physics (CCEPP); Joint Large-Scale Scientific Facility Funds of the NSFC and CAS under Contract No. U2032104, U1832207; CAS under Contract No. YSBR-101; 100 Talents Program of CAS; The Excellent Youth Foundation of Henan Scientific Committee under Contract No.~242300421044; The Institute of Nuclear and Particle Physics (INPAC) and Shanghai Key Laboratory for Particle Physics and Cosmology; Agencia Nacional de Investigación y Desarrollo de Chile (ANID), Chile under Contract No. ANID PIA/APOYO AFB230003; German Research Foundation DFG under Contract No. FOR5327; Istituto Nazionale di Fisica Nucleare, Italy; Knut and Alice Wallenberg Foundation under Contracts Nos. 2021.0174, 2021.0299; Ministry of Development of Turkey under Contract No. DPT2006K-120470; National Research Foundation of Korea under Contract No. NRF-2022R1A2C1092335; National Science and Technology fund of Mongolia; National Science Research and Innovation Fund (NSRF) via the Program Management Unit for Human Resources \& Institutional Development, Research and Innovation of Thailand under Contract No. B50G670107; Polish National Science Centre under Contract No. 2019/35/O/ST2/02907; Swedish Research Council under Contract No. 2019.04595; The Swedish Foundation for International Cooperation in Research and Higher Education under Contract No. CH2018-7756; U. S. Department of Energy under Contract No. DE-FG02-05ER41374

\onecolumngrid

\begin{appendices}

\section{End Matter}

\subsection{Projections on helicity angles}~\label{sec:proj_hel_angle}

\begin{figure*}[htbp]
	\centering
	\centering
	\setlength{\abovecaptionskip}{0.cm}
	\includegraphics[width=14cm]{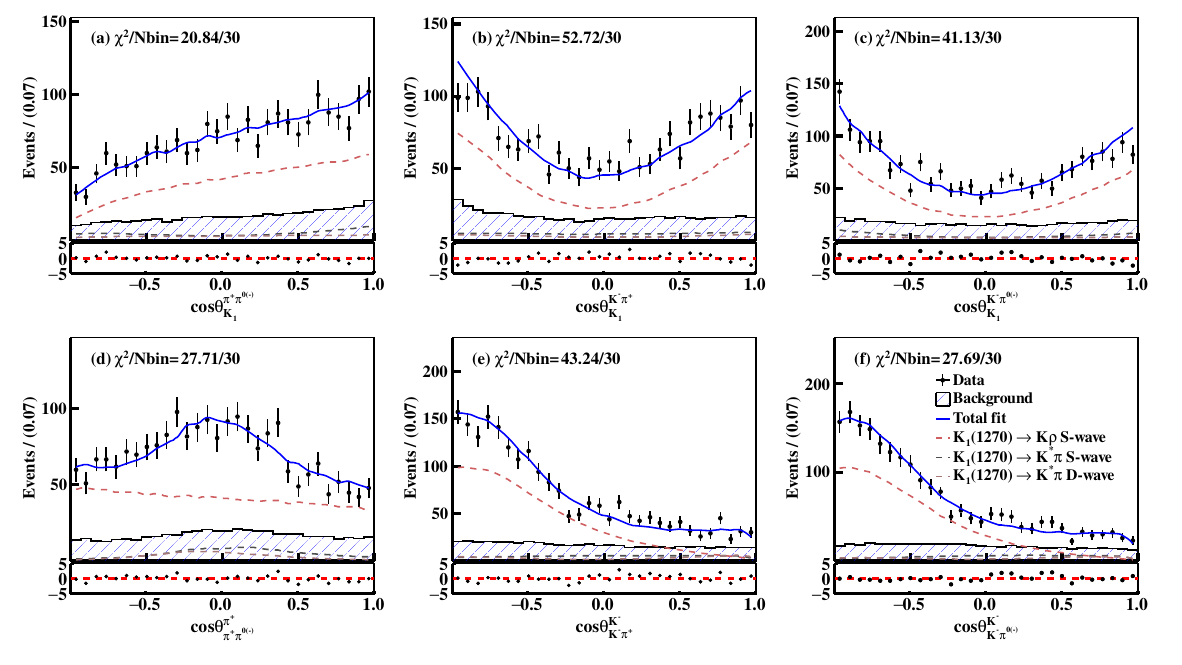}
	\caption{Projections of the amplitude analysis on the helicity angles (a) $\cos\theta_{K_1}^{\pi^+\pi^{0(-)}}$, (b) $\cos\theta_{K_1}^{K^-\pi^{+}}$, (c) $\cos\theta_{K_1}^{K^-\pi^{0(-)}}$, (d) $\cos\theta_{\pi^+\pi^{0(-)}}^{\pi^+}$, (e) $\cos\theta_{K^-\pi^+}^{K^-}$, and (f) $\cos\theta_{K^-\pi^{0(-)}}^{K^-}$. The dots with error bars are data, the blue solid curves are the total fits, the dashed curves are the various components, and the blue hatched histograms are the backgrounds. The pull distributions defined as $\chi=(N_{\rm data}-N_{\rm fit})/\sqrt{N_{\rm fit}}$ are shown in the lower plots, where $N_{\rm data}$ and $N_{\rm fit}$ are the data and fit number of events of projections for each bin, respectively.}
	\label{fig:amp_nominal_helicity}
\end{figure*}

The projections of the amplitude analysis on the helicity angles in the secondary decay $\bar{K}_1(1270)\to K^-\pi^+\pi^{0(-)}$ are shown in Fig.~\ref{fig:amp_nominal_helicity}.

\subsection{Angular fit on $\cos\theta_{L}$ v.s. $\cos\theta_K$}~\label{sec:angular_fit}

\begin{figure*}
	\centering
	\includegraphics[width=14 cm]{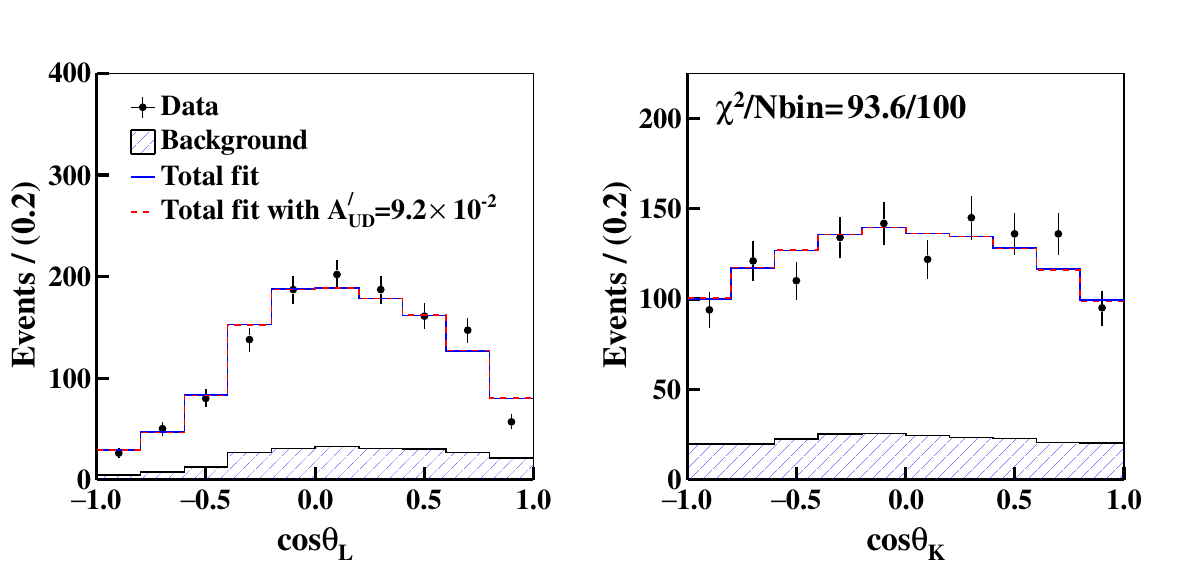}
	\caption{Fit projections of angular fit on $\cos\theta_l$~(left) and $\cos\theta_{K}$~(right). The blue curves are the fit results with float $\mathcal{A}^\prime_{ud}$. The red curves are the fit results with the predicted value $\mathcal{A}^\prime_{ud}=0.092$ as a reference.}
	\label{fig:angular_fit}
\end{figure*}

The distribution function $f(\cos\theta_{L},\cos\theta_K; \mathcal{A}^\prime_{ud}, d_{+}, d_{-})$ cited from Ref.~\cite{Fan:2021mwp} is written as
\begin{equation}
\begin{split}
&f(\cos\theta_{K},\cos\theta_{l};\mathcal{A}^\prime_{ud},d_{+},d_{-})=(4+d_{+}+d_{-})[1+\cos^2\theta_{K}\cos^2\theta_{l}]\\
&+2(d_{+}-d_{-})[1+\cos^2\theta_{K}]\cos\theta_{l}+2\mathcal{A}^\prime_{ud}(d_{+}-d_{-})[1+\cos^2\theta_{l}]\cos\theta_{K}\\
&+4\mathcal{A}^\prime_{ud}(d_{+}+d_{-})\cos\theta_{K}\cos\theta_{l}-(4-d_{+}-d_{-})[\cos^2\theta_{K}+\cos^2\theta_{l}].\\
\end{split}
\end{equation}
A binned 2D $\chi^2$ fit is  performed to extract $\mathcal{A}^\prime_{ud}$, $d_+$, and $d_-$.  The bin division is $10\times10$ in the range of $\cos\theta_{l}\in[-1,1]\&\cos\theta_{K}\in[-1,1]$. The $\chi^2$ is defined as
\begin{equation}
\chi^2=\sum_{i,j} (N^{\rm data}_{ij}-(N^{\rm bkg}_{ij}+N^{\rm sig}_{ij}))^2/(N^{\rm bkg}_{ij}+N^{\rm sig}_{ij}).
\end{equation}
Here, the $N^{\rm data}_{ij}$ and $N^{\rm bkg}_{ij}$ are the number of events of the data and the normalized inclusive MC sample in bin $(i,j)$. The $N^{\rm sig}_{ij}$ is the expected signal number with efficiency considered as
\begin{equation}
N^{\rm sig}_{ij}(A^\prime_{UD},d_{+},d_{-}) = \frac{\sum_{\alpha}^{N^{\rm un}_{ij}} \epsilon_\alpha\times f_{\alpha}(\cos\theta_{l},\cos\theta_{K};A^\prime_{UD},d_{+},d_{-})}{{\sum_{\alpha}^{N^{\rm un}} \epsilon_\alpha \times f_{\alpha}(\cos\theta_{l},\cos\theta_{K};A^\prime_{UD},d_{+},d_{-})}}\times(N^{\rm data}-N^{\rm bkg}).
\end{equation}
Here, $\epsilon_\alpha$ is the binned efficiency determined with signal MC sample as $\epsilon_{\alpha\in(i,j)}=N^{\rm rec}_{ij}/N^{\rm tru}_{ij}$ to take into account detector effects. The $N^{\rm un}=$ 100,000 is the number of random sets $(\cos\theta_{l},\cos\theta_{K})$ generated uniformly in the range of $\cos\theta_{l}\in[-1,1]\&\cos\theta_{K}\in[-1,1]$ and $N^{\rm un}_{ij}$ indicates the set falls into the bin $ij$. The simultaneous fit is performed by minimizing $\chi^2_{\rm sum}=\chi^2_{D^+}+\chi^2_{D^0}$. The fit projections are shown in Fig.~\ref{fig:angular_fit}.

\section{Supplemental Material}

\subsection{Single and double Tag information}

\begin{figure}[htbp]
	\centering
	\includegraphics[width=12.0cm]{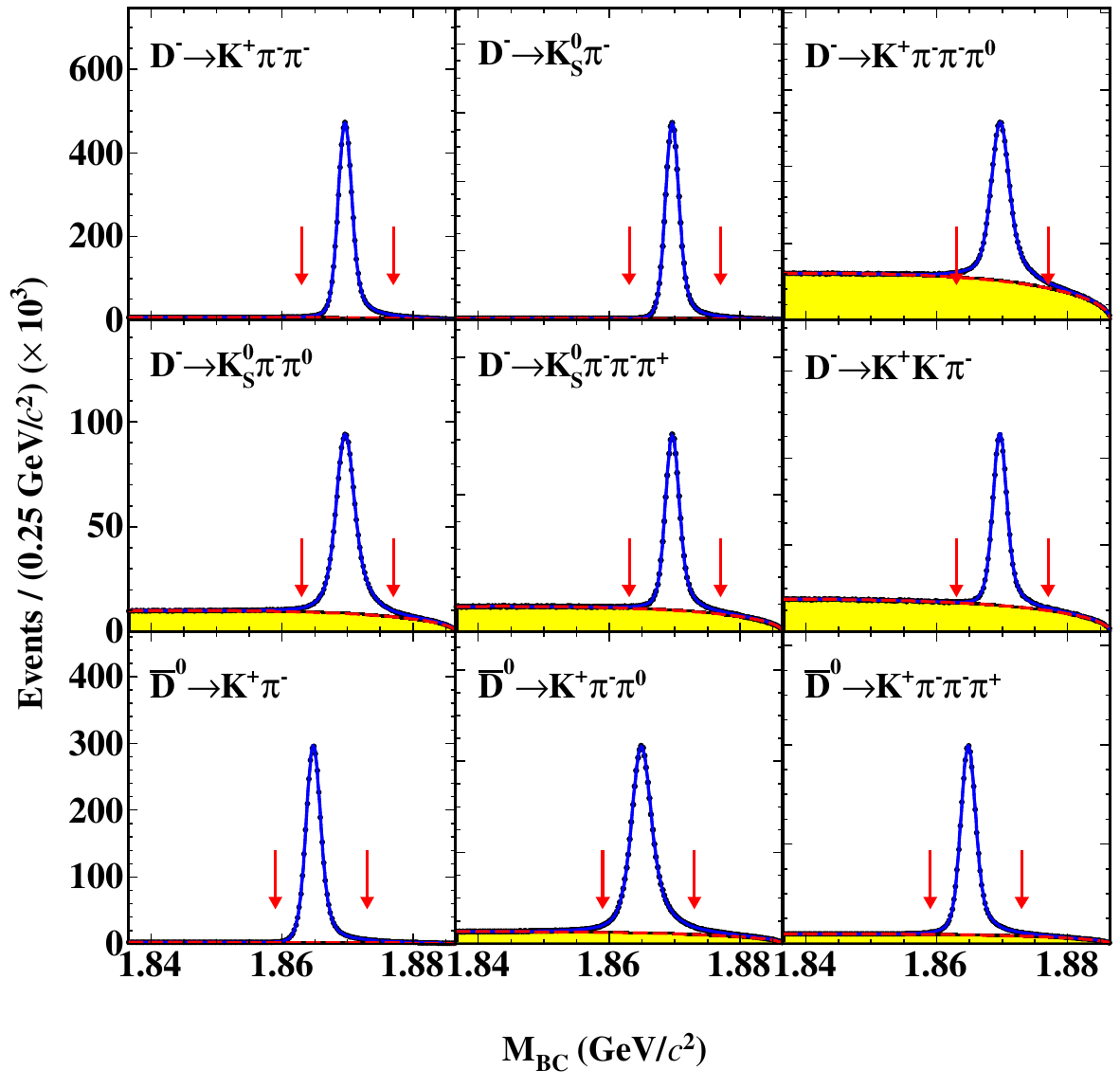}
	\caption{Fits to the $M_{\rm BC}$ distributions of tag channels. The dots with error bars are data, the blue solid curves are the total fits, the red dashed lines indicate the background contributions. The pair of red arrows indicate the $M_{\rm BC}$ signal window.}
	\label{fig:STfit}
\end{figure}

Figure~\ref{fig:STfit} shows the fits to the $M_{\rm BC}$ distributions of the selected ST candidates in data for different tag modes. Table~\ref{tab:STyield} summaries the requirements on $\Delta E$, the ST yields in data $N^{i}_{\rm ST}$, the ST efficiencies $\epsilon^{i}_{\rm ST}$, the DT efficiencies $\epsilon^{i}_{\rm DT}$, and the signal efficiencies $\epsilon^{i}_{\rm sig}=\epsilon^{i}_{\rm DT}/\epsilon^{i}_{\rm ST}$. 


\begin{table}[H]
	\centering
	\caption{The $\Delta E$ requirements, ST $\bar{D}$ yields in data ($N^{i}_{\rm ST}$), the ST efficiencies ($\epsilon^{i}_{\rm ST}$), the DT efficiencies ($\epsilon^{i}_{\rm DT}$), and the signal efficiencies ($\epsilon^{i}_{\rm sig}$).}
	\begin{tabular}{lccccc}
		\hline\hline
		Tag mode & $\Delta E$(GeV)  &  $N^{i}_{\rm ST}$~($\times 10^{4}$)     &  $\epsilon^{i}_{\rm ST} (\%)$ & $\epsilon^{i}_{\rm DT}$ (\%)&  $\epsilon^{i}_{\rm sig} (\%)$ \\\hline
		$\bar{D}^{0} \to K^{+} \pi^{-}$ &  $(-0.027,0.027)$  & $372.6 \pm 0.2$ & $65.10 \pm 0.01$ & $10.15\pm0.04$ & $15.58\pm0.05$ \\
		$\bar{D}^{0} \to K^{+} \pi^{-} \pi^{0}$ &  $(-0.062,0.049)$ & $743.3 \pm 0.3$ & $35.41 \pm 0.01$ & $5.13\pm0.01$ & $14.49\pm0.04$\\
		$\bar{D}^{0} \to K^{+} \pi^{+} \pi^{-} \pi^{-}$ &  $(-0.026,0.024)$  & $498.8 \pm 0.3$ & $40.96 \pm 0.01$ & $5.11\pm0.02$ & $12.48\pm0.04$ \\
		\hline
		Sum &     & $1614.6 \pm 0.5$ & & & \\
		\hline
		$D^{-} \to K^{+} \pi^{-} \pi^{-}$ &  $(-0.025,0.024)$ &  $556.7 \pm 0.3$ & $51.08 \pm 0.01$ & $5.06\pm0.02$ & $9.91\pm0.03$ \\
		$D^{-} \to K_{S}^{0} \pi^{-}$ &  $(-0.025,0.026)$ & $65.7 \pm 0.1$ & $51.42 \pm 0.01$ & $5.21\pm0.05$ & $10.13\pm0.09$\\
		$D^{-} \to K^{+} \pi^{-} \pi^{-} \pi^{0}$ &  $(-0.057,0.046)$  & $174.0 \pm 0.2$ & $24.53 \pm 0.01$ & $2.10\pm0.01$ & $8.56\pm0.05$\\
		$D^{-} \to K_{S}^{0} \pi^{-} \pi^{0}$ &  $(-0.062,0.049)$ & $144.2 \pm 0.2$ & $26.45 \pm 0.01$ & $2.43\pm0.02$ & $9.20\pm0.06$\\
		$D^{-} \to K_{S}^{0} \pi^{-} \pi^{-} \pi^{+}$ &  $(-0.028,0.027)$ & $79.0 \pm0.1$ & $29.68 \pm 0.01$ & $2.51\pm0.02$ & $8.36\pm0.08$\\
		$D^{-} \to K^{+} K^{-} \pi^{-}$ &  $(-0.024,0.023)$  & $48.1 \pm 0.1$ & $40.91 \pm 0.01$ & $3.70\pm0.04$ & $9.04\pm0.11$\\
		\hline
		Sum &   &   $1067.8 \pm 0.4$ & &  &\\
		\hline\hline
	\end{tabular}
	
	\label{tab:STyield}
\end{table}

\subsection{Amplitude formulas}

The covariant tensor amplitude of $D\to \bar{K}_1 e^+\nu$,$\bar{K}_1 \to K^-\pi\pi$ is constructed as described below.

The weak decay of $D\to \bar{K}_1 W^*$ is described with the $(V-A)$ current following the convention in Ref.~\cite{Bian:2021gwf} as
\begin{equation}
	\begin{aligned}
		(V-A)^{\mu\eta}\epsilon^{*}(\lambda_{W})_{\mu}\epsilon^{*}(\lambda_{K_1})_{\eta},
	\end{aligned}
\end{equation}
where $\epsilon(\lambda_{W/\bar{K}_1})$ is the polarization vector of $W^*$ and $\bar{K}_1$. The vector and axial vector currents are given as
\begin{equation}
	\begin{aligned}
		V^{\mu} &= V^{\mu\eta} \epsilon^{*}(\lambda_{K_1})_{\eta} 
		= \left(-2 M_{K_1} V_0(q^2) \frac{q^\eta q^\mu}{q^2} -(m_D-M_{K_1})V_1(q^2)\left[g^{\mu\eta}-\frac{q^\eta q^\mu}{q^2} \right] \right.\\
		&\left.+V_2(q^2) (\frac{q^\eta}{m_D-M_{K_1}})\left[(p_D+p_{K_1})^\mu-\frac{m_D^2-M_{K_1}^2}{q^2} q^\mu\right]\right)\epsilon^{*}(\lambda_{K_1})_{\eta}, \\
		A^{\mu} &= A^{\mu\eta} \epsilon^{*}(\lambda_{K_1})_{\eta} 
		= \left(-\frac{2 i A(q^2)}{m_D-M_{K_1}} \epsilon^{\mu \eta p^{D}p^{K_1}}\right)\epsilon^{*}(\lambda_{K_1})_{\eta},
	\end{aligned}
\end{equation}
respectively. Here $q^\mu=p^\mu_{D}-p_{K_1}^\mu$ is the momentum transfer, and $V_1,V_2,A$ are the form factors. In the lepton mass zero limit, the terms containing $q_\mu$ do not contribute to the amplitude with $q^\mu L_{\mu}=0$~\cite{Korner:1989qb}, where $L_{\mu}=\bar{u}_\nu\gamma_{\mu}(1-\gamma_5)v_l$ is the leptonic current of $W^*\to e^+\nu_e$. Therefore,
\begin{equation}
	\begin{aligned}
		&A^{\mu\eta}=-\frac{2 i A(q^2)}{m_D-M_{K_1}} \epsilon^{\mu \eta p^{D}p^{K_1}}, \\
		&V^{\mu\eta} =
		-(m_D-M_{K_1}) V_1(q^2)g^{\mu\eta} +V_2(q^2) (\frac{q^\eta}{m_D-M_{K_1}})(p_D+p_{K_1})^\mu.
	\end{aligned}
\end{equation}
Both vector and axial-vector FFs take single pole form, which are written as  $V_{1,2}(q^2)= \frac{V_{1,2}(0)}{1-q^2/m_V^2}$ and $A(q^2)=\frac{A(0)}{1-q^2/m_A^2}$. Here, $m_A=2.5$ GeV/$c^2$ and $m_{V}=2.1$ GeV/$c^2$ are the dominant pole mass of axial vector and vector currents, respectively. The hadronic decay of $\bar{K}_1\to K^-\pi\pi$ is described with $\epsilon(\lambda_{K_1})_{\sigma} J^{\sigma}$. Additionally, the Breit-Wigner function $\mathcal{R}_{\bar{K}_1}=\frac{1}{s-m_0^2+im_0\Gamma_0}$ is included due to the non-zero width of $\bar{K}_1(1270)$. Combining the above equations and summing over the polarization of $W^*$ and $\bar{K}_1$, the total amplitude is written as
\begin{equation}
	\begin{aligned}
		A=(V-A)^{\mu\eta}\cdot
		[\sum_{\lambda_{W}}\epsilon^{*}(\lambda_{W})_{\mu}\epsilon(\lambda_{W})_{\rho}]\cdot
		[\sum_{\lambda_{K_1}}\epsilon^{*}(\lambda_{K_1})_{\eta}\epsilon(\lambda_{K_1})_{\sigma}]\cdot\mathcal{R}_{\bar{K}_1}\cdot J^{\sigma}\cdot L^\rho = H_{\rho}L^\rho,
	\end{aligned}
\end{equation}
where the hadronic part $H_{\rho}$ is given as
\begin{equation}
	\label{eq:hadronic_part}
	\begin{aligned}
		H_{\rho} =& \mathcal{R}_{\bar{K}_1}\cdot(V-A)^{\mu\eta}g_{\mu\rho}(g_{\eta\sigma}-\frac{p^{K_1}_{\eta}p^{K_1}_{\sigma}}{M^2_{K_1}})J^{\sigma} = \mathcal{R}_{\bar{K}_1}\cdot\left(\frac{2 i A(q^2)}{m_D-M_{K_1}}\epsilon^{\rho J p^D p^{K_1}} \right.\\
		&\left.- V_1(q^2)(M_{K_1}-m_{D})\left[\frac{(J\cdot p^{K_1})}{M^2_{K_1}}p_{K_1}^\rho - J^{\rho}\right] -  \frac{V_2(q^2)}{m_D-M_{K_1}}\left[\frac{(J\cdot p^{K_1})(p^{K_1}\cdot q)}{M^2_{K_1}}-(J\cdot q)\right](p_{K_1}+p_{D})^\rho\right).
	\end{aligned}
\end{equation}
Finally, the $|A|^2$ after summing over the spin of $e^+$ and $\nu_e$ is calculated as
\begin{equation}
	\begin{aligned}
		\sigma \propto& \sum_{\lambda_{l}}\sum_{\lambda_{\nu}} |A|^2 = \sum_{\lambda_{l}}\sum_{\lambda_{\nu}}\bar{v}_{l}(1+\gamma_5)\gamma^\mu u_{\nu}~\bar{u}_\nu\gamma^{\rho}(1-\gamma_5)v_l \cdot H^{*}_{\mu} H_{\rho}\\
		=&-2i\epsilon^{HH^{*}p^{l}p^{\nu}}+ 2(H\cdot p^{\nu})(H^*\cdot p^{l}) + 2(H\cdot p^{l})(H^*\cdot p^{\nu}) - 2(H\cdot H^*)(p^{l}\cdot p^{\nu}),
	\end{aligned}
\end{equation}
where $p^l$ and $p^\nu$ are the four-momenta of lepton and neutrino, respectively.

The current $J^{\mu}$ of decay $\bar{K}_1\to K^-\pi\pi$ is constructed following Ref.~\cite{Zou:2002ar}. Potential intermediate states $\bar{K}^*(892)\to K^-\pi$, $\rho(770)\to\pi\pi$, $\omega\to\pi\pi$, and ${({\bar{K}}\pi)_ {S-{\rm wave}}}\to \bar{K}\pi$ are considered. The obtained formulas of $\epsilon(\lambda_{K_1})_{\mu}J^{\mu}$ are shown in Table~\ref{tab:K1270_to_Kpipi_amp}, where $\epsilon^\prime(m^\prime)$ is the polarization vector of the intermediate state with polarization $m^\prime$. The coupling coefficients $g=|g|e^{i\phi}$ are constrained with isospin relationship. In the fit, the coupling coefficient of the largest contribution $K\rho(770)$~($S$-wave) is fixed to 1. For all the two body hadronic decays $a\to bc$, the Blatt-Weisskopf barrier factor $X_L(p)$ is considered in this work. The explicit expression is written as
\begin{equation}
	\begin{split}
		X_{L=0}(q) &= 1, \\
		X_{L=1}(q) &= \sqrt{\frac{z^2_0+1}{z^2+1}}, \\
		X_{L=2}(q) &= \sqrt{\frac{z^4_0+3z^2_0+9}{z^4+3z^2+9}},
	\end{split}
\end{equation}
where $z=qR$, $z_0=q_0R$ and the radius parameter $R$ is fixed to be $3.0$ GeV$^{-1}$. The $q$ is the momentum of $b(c)$ in $a$ rest frame and $q_0$ is the value at the nominal mass of $a$.

\begin{table}[htbp]
	\centering
	\scriptsize
	\setlength\tabcolsep{12pt}
	\caption{The covariant tensor amplitudes of all possible processes in the decays $\bar{K}^0(1270)\to \pi^+_{1}\pi^0_{2} K^-_{3}$ and $K_1^{-}(1270)\to \pi^+_1\pi^-_2K^-_3$ constructed following Ref.~\cite{Zou:2002ar}.}
	\begin{tabular}{l|c|c}
		\hline
		\hline
		Process in $\bar{K}^0(1270)\to \pi^+_{1}\pi^0_{2} K^-_{3}$  & Formula of $\epsilon(\lambda_{K_1})_{\mu}J^{\mu}$ & Coefficient\\
		\hline
		$\bar{K}^0_1\to \rho^{+}_{12} K^-_{3}$~($S$-wave), $\rho_{12}\to \pi^+_1 \pi^0_2$  & \multirow{3}{*}{$\sum_{m^\prime} \epsilon(\lambda_{K_1})_{\mu}\epsilon^{\prime *\mu}(m^\prime) \mathcal{R}(M) \epsilon^{\prime *\alpha}(m^\prime)\widetilde{t}_{\alpha}(q^\prime)$} & $g^{S}_{\rho}$\\
		$\bar{K}^0_1\to \pi^0_2 \bar{K}^{*0}_{13}$~($S$-wave), $\bar{K}^{*0}_{13}\to \pi^+_1 K^-_3$  & & $-g^{S}_{K^*}$ \\
		$\bar{K}^0_1\to \pi^+_1 {K}^{*-}_{23}$~($S$-wave), ${K}^{*-}_{23}\to \pi^0_2 K^-_3$  &  &$g^{S}_{K^*}$\\
		\hline
		$\bar{K}^0_1\to \rho^{+}_{12} K^-_{3}$~($D$-wave), $\rho_{12}\to \pi^+_1 \pi^0_2$ & \multirow{3}{*}{  $\sum_{m^\prime}\epsilon(\lambda_{K_1})^{\mu}\epsilon^{\prime *\nu}(m^\prime)\widetilde{t}_{\mu\nu}(q) \mathcal{R}(M) \epsilon^{\prime *\alpha}(m^\prime)\widetilde{t}_{\alpha}(q^\prime)$} & $g^{D}_{\rho}$  \\
		$\bar{K}_1^{0}\to \pi^0_2\bar{K}^{*0}_{13}$~($D$-wave), $\bar{K}^{*0}_{13}\to\pi^+_1 K^-_3$  & & $-g^{D}_{K^*}$ \\
		$\bar{K}_1^{0}\to \pi^+_1 {K}^{*-}_{23}$~($D$-wave), ${K}^{*-}_{23}\to\pi^0_2 K^-_3$  & & $g^{D}_{K^*}$\\
		\hline
		$\bar{K}_1^{0}\to \pi^0_2 {(\bar{K}\pi)_ {S-{\rm wave}}}_{13}$~($P$-wave), ${(\bar{K}\pi)_ {S-{\rm wave}}}_{13}\to\pi^+_1 K^-_3$ & \multirow{2}{*}{$\epsilon(\lambda_{K_1})_{\mu}\widetilde{t}^{\mu}(q)\mathcal{R}(M)$} & $-g^{P}_{K_0}$ \\
		$\bar{K}^0_1\to \pi^+_1 {(\bar{K}\pi)_ {S-{\rm wave}}}_{23}$~($P$-wave), ${(\bar{K}\pi)_ {S-{\rm wave}}}_{23}\to \pi^0_2 K^-_3$  &  &$g^{P}_{K_0}$\\
		\hline
		\hline
		Process in $K_1^{-}(1270)\to \pi^+_1\pi^-_2K^-_3$  & Formula & coefficient\\
		\hline
		$K_1^{-}\to \rho^{0}_{12} K^-_3$~($S$-wave), $\rho_{12}\to\pi^+_1 \pi^-_2$  & \multirow{3}{*}{$\sum_{m^\prime} \epsilon(\lambda_{K_1})_{\mu}\epsilon^{\prime *\mu}(m^\prime) \mathcal{R}(M) \epsilon^{\prime *\alpha}(m^\prime)\widetilde{t}_{\alpha}(q^\prime)$} & $-\frac{1}{\sqrt{2}} g^{S}_{\rho}$\\
		$K_1^{-}\to \pi^-_2\bar{K}^{*0}_{13}$~($S$-wave), $\bar{K}^{*0}_{13}\to\pi^+_1 K^-_3$  &  &$\sqrt{2} g^{S}_{K^*}$\\
		$K_1^{-}\to \omega^{0}_{12} K^-_3$~($S$-wave), $\omega_{12}\to\pi^+_1 \pi^-_2$ & & $g^{S}_{\omega}$\\
		\hline
		$K_1^{-}\to \rho^0_{12} K^-_{3}$~($D$-wave), $\rho^0_{12}\to\pi^+_1 \pi^-_2$ & \multirow{2}{*}{  $\sum_{m^\prime}\epsilon(\lambda_{K_1})^{\mu}\epsilon^{\prime *\nu}(m^\prime)\widetilde{t}_{\mu\nu}(q) \mathcal{R}(M) \epsilon^{\prime *\alpha}(m^\prime)\widetilde{t}_{\alpha}(q^\prime)$} & $-\frac{1}{\sqrt{2}}g^{D}_{\rho}$  \\
		$K_1^{-}\to \pi^-_2\bar{K}^{*0}_{13}$~($D$-wave), $\bar{K}^{*0}_{13}\to\pi^+_1 K^-_3$  & & $\sqrt{2}g^{D}_{K^*}$ \\
		\hline
		$K_1^{-}\to \pi^-_2{(\bar{K}\pi)_ {S-{\rm wave}}}_{13}$~($P$-wave), ${(\bar{K}\pi)_ {S-{\rm wave}}}_{13}\to\pi^+_1 K^-_3$ & $\epsilon(\lambda_{K_1})_{\mu}\widetilde{t}^{\mu}(q)\mathcal{R}(M)$ & $\sqrt{2} g^{P}_{K_0}$ \\
		\hline
		\hline
	\end{tabular}
	\label{tab:K1270_to_Kpipi_amp}
\end{table}

The lineshapes of intermediate states $\mathcal{R}(M)$ are listed below:
\begin{itemize}
	\item $\rho(770)$ is parameterized with a Gounaris-Sakurai model, which is given by
	\begin{equation}
		\frac{1+D\Gamma_0/m_0}{(m^2_0-m^2)+f(m)-im_0\Gamma(m)}.
	\end{equation}
	Here, mass dependent width is $\Gamma(m)=\Gamma_0(\frac{q}{q_0})^{2L+1}(\frac{m_0}{m})\left(\frac{B_L(q)}{B_L(q_0)}\right)^2$. The $f(m)$ and $D$ are defined as
	\begin{equation}
		\begin{split}
			&f(m) = \Gamma_0 \frac{m_0^2}{q^3_0}\left[q^2[h(m)-h(m_0)]+(m_0^2-m^2)q^2_0\frac{{\rm d}h}{{\rm d}m}|_{m_0}\right],\\
			&h(m) = \frac{2}{\pi}\frac{q}{m}\ln\left(\frac{m+2q}{2m_\pi}\right),\\
			&\frac{{\rm d}h}{{\rm d}m}|_{m_0} = h(m_0)\left[(8q^2_0)^{-1}-(2m^2_0)^{-1}\right]+(2\pi m^2_0)^{-1},\\
			&D = \frac{f(0)}{\Gamma_0m_0} = \frac{3}{\pi}\frac{m^2_\pi}{q^2_0}\ln\left(\frac{m_0+2q_0}{2m_\pi}\right)+\frac{m_0}{2\pi q_0}-\frac{m^2_\pi m_0}{\pi q^3_0}.
		\end{split}
	\end{equation}
	\item $\omega$ is described by a Breit-Wigner function with constant width $\frac{1}{s-m_0^2+im_0\Gamma_0}$.
	\item $K^*(892)$ is described with a mass-dependent width as $\frac{1}{s-m_0^2+im_0\Gamma(m)}$ with $\Gamma(m)=\Gamma_0(\frac{q}{q_0})^{2L+1}(\frac{m_0}{m})(\frac{B_L(q)}{B_L(q_0)})^2$. 
	\item The $(K\pi)_{S-{\rm wave}}$ is modeled by a parameterisation from scattering data as $A(m)=F\sin\delta_Fe^{i\delta_F}+R\sin\delta_Re^{i\delta_R}e^{i2\delta_F}$, with parameters fixed to Ref.~\cite{BaBar:2018cka}.
\end{itemize}

The construction of the amplitude of $\bar{K}_1(1400)\to K^-\pi\pi$ is the same as for the $\bar{K}_1(1270)$ contribution. To incorporate $\bar{K}_1(1400)$ into the amplitude in the significance test, the hadronic part $H_\rho$ in Eq.~\ref{eq:hadronic_part} sums over both contributions from $\bar{K}_1(1270)$ and $\bar{K}_1(1400)$. The mass and width of $\bar{K}_1(1400)$ are fixed to the individual PDG average values. 

\subsection{Systematic uncertainties for amplitude analysis}

Table~\ref{tab:sum_sys_amp} shows the systematic uncertainties for the amplitude analysis.

\begin{table}[htbp]
	\centering
	\setlength\tabcolsep{10pt}
	\caption{The systematic uncertainties for the fitted parameters and fit fractions in amplitude analysis.}
	\begin{tabular}{lccccccc}
		\hline
		\hline
		Variable & Insignificant & Fit bias & Radius & Lineshape & Background & FFs model & Sum \\ \hline
		$r_{A}~(\times10^{-2})$ & 0.66 & 0.54 & 0.10 & 0.01 & 0.05 & 0.08 & 0.86 \\
		$r_{V}~(\times10^{-2})$ & 2.27 & 0.86 & 0.02 & 0.05 & 0.05 & 0.51 & 2.48  \\
		$f^{D^+}_{\rho K^-}$~(\%)& 25.7 & 0.1 & 0.6 & 0.0 & 0.2 & 0.1 & 25.7  \\
		$f^{D^+}_{\pi \bar{K}^*(892)}$~(\%)& 2.9 & 0.4 & 0.2 & 0.2 & 0.1 &0.1 & 3.0  \\
		$f^{D^0}_{\rho K^-}$~(\%) & 23.9 & 0.2 & 0.5 & 0.1 & 0.1 & 0.1 & 23.9  \\
		$f^{D^0}_{\pi \bar{K}^*(892)}$~(\%)& 5.2 & 0.7 & 0.3 & 0.2 & 0.1 & 0.1 & 5.2  \\
		$m_{0}$ (MeV/$c^2$) & 4.7 & 3.3 & 3.8 & --- & 0.1 & 0.2 & 6.9  \\
		$\Gamma_{0}$ (MeV) & 18.0 & 4.5 & 6.6 & --- & 3.6 & 0.1 & 20.0  \\
		\hline
		\hline
	\end{tabular}
	\label{tab:sum_sys_amp}
\end{table}

\end{appendices}

\end{document}

%% file: authorlist_2024-11-29.tex
\author{
M.~Ablikim$^{1}$, M.~N.~Achasov$^{4,c}$, P.~Adlarson$^{76}$, X.~C.~Ai$^{81}$, R.~Aliberti$^{35}$, A.~Amoroso$^{75A,75C}$, Q.~An$^{72,58,a}$, Y.~Bai$^{57}$, O.~Bakina$^{36}$, Y.~Ban$^{46,h}$
% ...
\\
\vspace{0.2cm}
(BESIII Collaboration)\\
\vspace{0.2cm} {\it
$^{1}$ Institute of High Energy Physics, Beijing 100049, People's Republic of China\\
% ...
$^{4}$ Budker Institute of Nuclear Physics SB RAS (BINP), Novosibirsk 630090, Russia\\
% ...
$^{35}$ Johannes Gutenberg University of Mainz, Johann-Joachim-Becher-Weg 45, D-55099 Mainz, Germany\\
$^{36}$ Joint Institute for Nuclear Research, 141980 Dubna, Moscow region, Russia\\
% ...
$^{46}$ Peking University, Beijing 100871, People's Republic of China\\
% ...
$^{57}$ Southeast University, Nanjing 211100, People's Republic of China\\
$^{58}$ State Key Laboratory of Particle Detection and Electronics, Beijing 100049, Hefei 230026, People's Republic of China\\
% ...
$^{72}$ University of Science and Technology of China, Hefei 230026, People's Republic of China\\
% ...
$^{75}$ University of Turin and INFN, (A)University of Turin, I-10125, Turin, Italy; (B)University of Eastern Piedmont, I-15121, Alessandria, Italy; (C)INFN, I-10125, Turin, Italy\\
$^{76}$ Uppsala University, Box 516, SE-75120 Uppsala, Sweden\\
% ...
$^{81}$ Zhengzhou University, Zhengzhou 450001, People's Republic of China\\
\vspace{0.2cm}
$^{a}$ Deceased\\
% ...
$^{c}$ Also at the Novosibirsk State University, Novosibirsk, 630090, Russia\\
% ...
$^{h}$ Also at State Key Laboratory of Nuclear Physics and Technology, Peking University, Beijing 100871, People's Republic of China\\
% ...
}
}

%% file: draft_D2Kpipienu_v2.0.bbl
\begin{thebibliography}{99}


\bibitem{Cabibbo:1963yz}
N.~Cabibbo,
\href{https://doi.org/10.1103/PhysRevLett.10.531}{Phys. Rev. Lett. \textbf{10}, 531-533 (1963).}

\bibitem{Kobayashi:1973fv}
M.~Kobayashi and T.~Maskawa,
\href{https://doi.org/10.1143/PTP.49.652}{Prog. Theor. Phys. \textbf{49}, 652-657 (1973).}

\bibitem{FlavourLatticeAveragingGroupFLAG:2024oxs}
Y.~Aoki \textit{et al.} (Flavour Lattice Averaging Group),
\href{https://arxiv.org/pdf/2411.04268}{[arXiv:2411.04268 [hep-lat]]}.

\bibitem{Ke:2023qzc}
B.~C.~Ke, J.~Koponen, H.~B.~Li and Y.~Zheng,
\href{https://doi.org/10.1146/annurev-nucl-110222-044046}{Ann. Rev. Nucl. Part. Sci. \textbf{73}, 285-314 (2023).}

\bibitem{HFLAV:2022esi}
Y.~S.~Amhis \textit{et al.} (HFLAV Collaboration),
\href{https://doi.org/10.1103/PhysRevD.107.052008}{Phys. Rev. D \textbf{107}, no.5, 052008 (2023).}

\bibitem{Momeni:2019uag}
S.~Momeni and R.~Khosravi,
\href{https://doi.org/10.1088/1361-6471/ab35d0}{J. Phys. G \textbf{46}, 105006 (2019).}

\bibitem{Khosravi:2008jw}
R.~Khosravi, K.~Azizi and N.~Ghahramany,
\href{https://doi.org/10.1103/PhysRevD.79.036004}{Phys. Rev. D \textbf{79}, 036004 (2009).}

\bibitem{Momeni:2022gqb}
S.~Momeni and M.~Saghebfar,
\href{https://doi.org/10.1140/epjc/s10052-022-10413-x}{Eur. Phys. J. C \textbf{82}, 473 (2022).}

\bibitem{Cheng:2003sm}
H.~Y.~Cheng, C.~K.~Chua and C.~W.~Hwang,
\href{https://doi.org/10.1103/PhysRevD.69.074025}{Phys. Rev. D \textbf{69}, 074025 (2004).}

\bibitem{Verma:2011yw}
R.~C.~Verma,
\href{https://doi.org/10.1088/0954-3899/39/2/025005}{J. Phys. G \textbf{39}, 025005 (2012)}; H.~Y.~Cheng and X.~W.~Kang,
\href{https://doi.org/10.1140/epjc/s10052-017-5170-5}{Eur. Phys. J. C \textbf{77}, 587 (2017).}





\bibitem{Shi:2023kiy}
Y.~J.~Shi, J.~Zeng and Z.~F.~Deng,
\href{https://doi.org/10.1103/PhysRevD.109.016027}{Phys. Rev. D \textbf{109}, 016027 (2024).}

\bibitem{Suzuki:1993yc}
M.~Suzuki,
\href{https://doi.org/10.1103/PhysRevD.47.1252}{Phys. Rev. D \textbf{47}, 1252-1255 (1993).}

\bibitem{Divotgey:2013jba}
F.~Divotgey, L.~Olbrich and F.~Giacosa,
\href{https://doi.org/10.1140/epja/i2013-13135-3}{Eur. Phys. J. A \textbf{49}, 135 (2013).}

\bibitem{Hatanaka:2008xj}
H.~Hatanaka and K.~C.~Yang,
\href{https://doi.org/10.1103/PhysRevD.77.094023}{Phys. Rev. D \textbf{77}, 094023 (2008).}

\bibitem{Cheng:2011pb}
H.~Y.~Cheng,
\href{https://doi.org/10.1016/j.physletb.2011.12.013}{Phys. Lett. B \textbf{707}, 116-120 (2012).}

\bibitem{Blundell:1995au}
H.~G.~Blundell, S.~Godfrey and B.~Phelps,
\href{https://doi.org/10.1103/PhysRevD.53.3712}{Phys. Rev. D \textbf{53}, 3712-3722 (1996).}

\bibitem{Tayduganov:2011ui}
A.~Tayduganov, E.~Kou and A.~Le Yaouanc,
\href{https://doi.org/10.1103/PhysRevD.85.074011}{Phys. Rev. D \textbf{85}, 074011 (2012).}

\bibitem{Lipkin:1977uy}
H.~J.~Lipkin,
\href{https://doi.org/10.1016/0370-2693(77)90714-6}{Phys. Lett. B \textbf{72}, 249-250 (1977).}

\bibitem{Burakovsky:1997dd}
L.~Burakovsky and J.~T.~Goldman,
\href{https://doi.org/10.1103/PhysRevD.56.R1368}{Phys. Rev. D \textbf{56}, R1368-R1372 (1997).}

\bibitem{Cheng:2004yj}
H.~Y.~Cheng and C.~K.~Chua,
\href{https://doi.org/10.1103/PhysRevD.69.094007}{Phys. Rev. D \textbf{69}, 094007 (2004).}

\bibitem{Cheng:2003bn}
H.~Y.~Cheng,
\href{https://doi.org/10.1103/PhysRevD.67.094007}{Phys. Rev. D \textbf{67}, 094007 (2003).}


\bibitem{Guo:2018orw}
P.~F.~Guo, D.~Wang and F.~S.~Yu,
\href{https://doi.org/10.11804/NuclPhysRev.36.02.125}{Nucl. Phys. Rev. \textbf{36}, 125-134 (2019).}




\bibitem{Wang:2019wee}
W.~Wang, F.~S.~Yu and Z.~X.~Zhao,
\href{https://doi.org/10.1103/PhysRevLett.125.051802}{Phys. Rev. Lett. \textbf{125}, 051802 (2020).}

\bibitem{Bian:2021gwf}
L.~Bian, L.~Sun and W.~Wang,
\href{https://doi.org/10.1103/PhysRevD.104.053003}{Phys. Rev. D \textbf{104}, 053003 (2021).}

\bibitem{Atwood:1997zr}
D.~Atwood, M.~Gronau and A.~Soni,
\href{https://doi.org/10.1103/PhysRevLett.79.185}{Phys. Rev. Lett. \textbf{79}, 185-188 (1997).}

\bibitem{Becirevic:2012dx}
D.~Becirevic, E.~Kou, A.~Le Yaouanc and A.~Tayduganov,
\href{https://doi.org/10.1007/JHEP08(2012)090}{JHEP \textbf{08}, 090 (2012).}

\bibitem{Paul:2016urs}
A.~Paul and D.~M.~Straub,
\href{https://doi.org/10.1007/JHEP04(2017)027}{JHEP \textbf{04}, 027 (2017).}

\bibitem{LHCb:2014vnw}
R.~Aaij \textit{et al.} (LHCb Collaboration),
\href{https://doi.org/10.1103/PhysRevLett.112.161801}{Phys. Rev. Lett. \textbf{112}, 161801 (2014).}


\bibitem{CLEO:2007oer}
M.~Artuso \textit{et al.} (CLEO Collaboration),
\href{https://doi.org/10.1103/PhysRevLett.99.191801}{Phys. Rev. Lett. \textbf{99}, 191801 (2007).}

\bibitem{BESIII:2019eao}
M.~Ablikim \textit{et al.} (BESIII Collaboration),
\href{https://doi.org/10.1103/PhysRevLett.123.231801}{Phys. Rev. Lett. \textbf{123}, 231801 (2019).}

\bibitem{BESIII:2021uqr}
M.~Ablikim \textit{et al.} (BESIII Collaboration),
\href{https://doi.org/10.1103/PhysRevLett.127.131801}{Phys. Rev. Lett. \textbf{127}, 131801 (2021).}

\bibitem{BESIII:2024ieo}
M.~Ablikim \textit{et al.} (BESIII Collaboration),
\href{https://doi.org/10.1007/JHEP09(2024)089}{JHEP \textbf{09}, 089 (2024).}

\bibitem{BESIII:2024lbn}
M.~Ablikim \textit{et al.} (BESIII Collaboration),
\href{https://doi.org/10.1088/1674-1137/ad70a0}{Chin. Phys. C \textbf{48}, 123001 (2024).}

\bibitem{cc_ref}
Throughout this Letter, the charged-conjugation modes are always implied.


\bibitem{BESIII:2009fln}
M.~Ablikim \textit{et al.} (BESIII Collaboration),
\href{https://doi.org/10.1016/j.nima.2009.12.050}{Nucl. Instrum. Meth. A \textbf{614}, 345-399 (2010).}

\bibitem{Huang:2022wuo}
K.~X.~Huang, Z.~J.~Li, Z.~Qian, J.~Zhu, H.~Y.~Li, Y.~M.~Zhang, S.~S.~Sun and Z.~Y.~You,
\href{https://doi.org/10.1007/s41365-022-01133-8}{Nucl. Sci. Tech. \textbf{33}, 142 (2022).}

\bibitem{GEANT4:2002zbu}
S.~Agostinelli \textit{et al.} (GEANT4 Collaboration),
\href{https://doi.org/10.1016/S0168-9002(03)01368-8}{Nucl. Instrum. Meth. A \textbf{506}, 250-303 (2003).}

\bibitem{Jadach:1999vf}
S.~Jadach, B.~F.~L.~Ward and Z.~Was,
\href{https://doi.org/10.1016/S0010-4655(00)00048-5}{Comput. Phys. Commun. \textbf{130}, 260-325 (2000).}

\bibitem{Jadach:2000ir}
S.~Jadach, B.~F.~L.~Ward and Z.~Was,
\href{https://doi.org/10.1103/PhysRevD.63.113009}{Phys. Rev. D \textbf{63}, 113009 (2001).}

\bibitem{Lange:2001uf}
D.~J.~Lange,
\href{https://doi.org/10.1016/S0168-9002(01)00089-4}{Nucl. Instrum. Meth. A \textbf{462}, 152-155 (2001).}

\bibitem{Ping:2008zz}
R.~G.~Ping,
\href{https://doi.org/10.1088/1674-1137/32/8/001}{Chin. Phys. C \textbf{32}, 599 (2008).}

\bibitem{Richter-Was:1992hxq}
E.~Richter-Was,
\href{https://doi.org/10.1016/0370-2693(93)90062-M}{Phys. Lett. B \textbf{303}, 163-169 (1993).}


\bibitem{BESIII:2023exq}
M.~Ablikim \textit{et al.} (BESIII Collaboration),
\href{https://doi.org/10.1103/PhysRevD.109.072003}{Phys. Rev. D \textbf{109}, 072003 (2024).}

\bibitem{supplemental_material}
See Supplemental Material at [URL will be inserted by publisher] for ST yields and efficiencies, amplitude formulas, and systematic uncertainties for amplitude analysis.

\bibitem{BESIII:2015hty}
M.~Ablikim \textit{et al.} (BESIII Collaboration),
\href{https://doi.org/10.1103/PhysRevD.94.032001}{Phys. Rev. D \textbf{94}, 032001 (2016).}

\bibitem{ParticleDataGroup:2024cfk}
S.~Navas \textit{et al.} (Particle Data Group),
\href{https://doi.org/10.1103/PhysRevD.110.030001}{Phys. Rev. D \textbf{110}, 030001 (2024).}


\bibitem{BESIII:2017jyh}
M.~Ablikim \textit{et al.} (BESIII Collaboration),
\href{https://doi.org/10.1103/PhysRevD.95.072010}{Phys. Rev. D \textbf{95}, 072010 (2017)}.






\bibitem{Belle:2010wrf}
H.~Guler \textit{et al.} (Belle Collaboration),
\href{https://doi.org/10.1103/PhysRevD.83.032005}{Phys. Rev. D \textbf{83}, 032005 (2011).}

\bibitem{LHCb:2024cwp}
R.~Aaij \textit{et al.} (LHCb Collaboration),
\href{https://doi.org/10.1007/JHEP01(2025)054}{JHEP \textbf{01}, 054 (2025).}

\bibitem{LHCb:2017swu}
R.~Aaij \textit{et al.} (LHCb Collaboration),
\href{https://doi.org/10.1140/epjc/s10052-018-5758-4}{Eur. Phys. J. C \textbf{78}, 443 (2018).}

\bibitem{CLEO:2009svp}
D.~Besson \textit{et al.} (CLEO Collaboration),
\href{https://doi.org/10.1103/PhysRevD.80.032005}{Phys. Rev. D \textbf{80}, 032005 (2009).}

\bibitem{BESIII:2024slx}
M.~Ablikim \textit{et al.} (BESIII Collaboration),
\href{https://doi.org/10.1103/10.1103/PhysRevD.110.112006}{Phys. Rev. D \textbf{110}, 112006 (2024).}

\bibitem{BESIII:2021qfo}
M.~Ablikim \textit{et al.} (BESIII Collaboration),
\href{https://doi.org/10.1103/PhysRevD.104.032011}{Phys. Rev. D \textbf{104}, 032011 (2021)}

\bibitem{LHCb:2018mzv}
R.~Aaij \textit{et al.} (LHCb Collaboration),
\href{https://doi.org/10.1007/JHEP02(2019)126}{JHEP \textbf{02}, 126 (2019)}


\bibitem{Fan:2021mwp}
Y.~L.~Fan, X.~D.~Shi, X.~R.~Zhou and L.~Sun,
\href{https://doi.org/10.1140/epjc/s10052-021-09841-y}{Eur. Phys. J. C \textbf{81}, 1068 (2021)}.

\bibitem{BESIII:2024zfv}
M.~Ablikim \textit{et al.} (BESIII Collaboration),
\href{https://doi.org/10.1007/JHEP12(2024)206}{JHEP \textbf{12}, 206 (2024)}.



\bibitem{BESIII:2015tql}
M.~Ablikim \textit{et al.} (BESIII Collaboration),
\href{https://doi.org/10.1103/PhysRevD.92.072012}{Phys. Rev. D \textbf{92}, 072012 (2015).}


\bibitem{Belle-II:2022cgf}
L.~Aggarwal~\textit{et al.} (Belle-II Collaboration),
\href{https://arxiv.org/abs/2207.06307}{arXiv:2207.06307 [hep-ex].}

\bibitem{LHCb:2023hlw}
R.~Aaij \textit{et al.} (LHCb Collaboration),
\href{https://doi.org/10.1088/1748-0221/19/05/P05065}{JINST \textbf{19}, no.05, P05065 (2024).}

\bibitem{Cheng:2022tog}
H.~Y.~Cheng, X.~R.~Lyu and Z.~Z.~Xing,
\href{https://doi.org/10.1088/0256-307X/42/1/010201}{Chin. Phys. Lett. \textbf{42}, 010201 (2025).}

\end{thebibliography}

\begin{thebibliography}{99}
	
	\bibitem{Bian:2021gwf}
	L.~Bian, L.~Sun and W.~Wang,
	\href{https://doi.org/10.1103/PhysRevD.104.053003}{Phys. Rev. D \textbf{104}, no.5, 053003 (2021).}
	
	\bibitem{Korner:1989qb}
	J.~G.~Korner and G.~A.~Schuler,
	\href{https://doi.org/10.1007/BF02440838}{Z. Phys. C \textbf{46}, 93 (1990).}
	
	\bibitem{Zou:2002ar}
	B.~S.~Zou and D.~V.~Bugg,
	\href{https://doi.org/10.1140/epja/i2002-10135-4}{Eur. Phys. J. A \textbf{16}, 537-547 (2003).}
	
	\bibitem{BaBar:2018cka}
	I.~Adachi \textit{et al.} (BaBar and Belle Collaborations),
	\href{https://doi.org/10.1103/PhysRevD.98.112012}{Phys. Rev. D \textbf{98}, no.11, 112012 (2018)}.
	
	\bibitem{Fan:2021mwp}
	Y.~L.~Fan, X.~D.~Shi, X.~R.~Zhou and L.~Sun,
	\href{https://doi.org/10.1140/epjc/s10052-021-09841-y}{Eur. Phys. J. C \textbf{81}, no.12, 1068 (2021)}.
	
\end{thebibliography}
